\documentclass{ws-procs975x65}
\makeindex
\usepackage{hyperref}
\hypersetup{
    unicode=false,          % non-Latin characters in Acrobat
    colorlinks=true,       % false: boxed links; true: colored links
    linkcolor=black,          % color of internal links
    citecolor=black,        % color of links to bibliography
    filecolor=black,      % color of file links
    urlcolor=black           % color of external links
}
\usepackage{enumerate}

\def\be {\begin{equation}}
\def\ee {\end{equation}}
\def\beq{\begin{equation}}
\def\eeq{\end{equation}}
\def\bea {\begin{eqnarray}}
\def\eea {\end{eqnarray}}
\def\br{\begin{eqnarray}}
\def\er{\end{eqnarray}}
\def\bc {\begin{center}}
\def\ec {\end{center}}
\def\bfg {\begin{figure}}
\def\efg {\end{figure}}
\def\bi {\begin{itemize}}
\def\ei {\end{itemize}}
\def\benu{\begin{enumerate}}
\def\eenu{\end{enumerate}}
\newcommand{\bdm}{\begin{displaymath}}
\newcommand{\edm}{\end{displaymath}}

\def\l{\left}
\def\r{\right}

\def\laq{\hbox{~}\raise 0.4ex\hbox{$<$}\kern -0.8em\lower 0.62ex\hbox{$\sim$}\hbox{~}}
\def\gaq{\hbox{~}\raise 0.4ex\hbox{$>$}\kern -0.7em\lower 0.62ex\hbox{$\sim$}\hbox{~}}

\def\MPl{M_{_{\rm Pl}}^2}

\newsavebox{\blambox}\savebox{\blambox}[0.6em]{$\lambda\!\!\!$\raisebox{0.5em}
{$\neg$}}\newcommand{\blambda}{\usebox{\blambox}}
\newsavebox{\bFox}\savebox{\bFox}[0.6em]{$F\!\!\!\!$\raisebox{0.5em}
{$\neg$}}
\newsavebox{\bxibox}\savebox{\bxibox}[0.6em]{$\xi\!\!\!$\raisebox{.5em}
{$\neg$}}

\newcommand{\bphi}{\overline{\varphi}}
\newcommand{\dphi}{\delta \varphi}
\newcommand{\Hm}{\mathcal{H}}

\newcommand{\Fpt}{{\mathcal{F}}}
\newcommand{\Fpteps}{{\mathcal{F}}_{\varepsilon}}

\begin{document}
\title{Dark spinor driven inflation}

\author{S.~Shankaranarayanan}

\address{${~}^{(1)}$ Institute of Cosmology and Gravitation, University of Portsmouth, Portsmouth, UK \\
${~}^{(2)}$ School of Physics, Indian Institute of Science Education and Research-Trivandrum, \\
CET campus, Thiruvananthapuram 695 016, India~~~
$^*$E-mail: shanki@iisertvm.ac.in}

\begin{abstract}
Inflation is considered to be the best paradigm
for describing the early universe. However, it is still
unclear what is the nature of the field which drives inflation. In this
talk, we discuss the possibility of spinor field driving inflation.
Spinflaton -- a scalar condensate of the dark spinor field -- has a single
scalar degree of freedom and leads to the identical acceleration
equation as the scalar field. We also discuss the advantages of
this model compared to the scalar field driven inflation and discuss its
observational relevance.
\end{abstract}

\bodymatter
%%%%%%%%%%%%%%%%%%
\vspace*{-0.005pt}
\section{Introduction}

Predictions of inflation seem to be in agreement with the CMB data \cite{Komatsu2008}. 
It is usually assumed that {\it inflaton} is an elementary
scalar field.  Based on this assumption and
field evolves slowly, inflation predicts: (i) the power
spectrum of the density fluctuations is almost scale invariant and has
no running. 
%(ii) that existence of background primordial gravitational waves and 
(ii) tensor to scalar ratio $(r) = 16
\varepsilon_{\rm can}$ where $\varepsilon_{\rm can}$ is slow-roll
parameter. It has been argued that such a relation, if observationally
verified, would offer strong support for the idea of inflation.

In this talk, we critically analyze this claim by considering a model
in which the inflation is not an elementary field. More precisely, we
ask the following question: If the inflaton is not an elementary
field, how robust are these predictions? It is long known that the
role played by inflaton can also be played by the curvature scalar $R$
%\cite{Starobinsky1980}, 
or logarithm of the radius of compactified
space 
%\cite{Shafi:1986vv} 
or vector meson condensate 
%\cite{Ford1989}
or a fermionic condensate. Although there has been intense activity in
several of these cases recently \cite{Ford1989}, the possibility of
fermionic condensate has not been discussed much in the literature
\cite{Boehmer2007i}. We show that a spinor condensate is a viable
alternative model for scalar driven inflation. In FRW space-time, the spinor
condensate has identical acceleration equation while the Friedman
equation is modified. We show that 
tensor and scalar spectra have running and satisfy different consistency 
relations.

Recently, the 
field theory for the eigen spinors of charge conjugation
operator (Majorana spinors) were constructed by Ahluwalia-Khalilova
and Grumiller \cite{Ahluwalia2005} and referred them
as Elkos. They showed that these new spinors possess special
properties under discrete symmetries like Charge ${\cal C}$, parity
${\cal P}$ and Time ${\cal T}$ operators. More precisely, they showed
that ${\cal P}^2 = -1, [{\cal C}, {\cal P}] = 0, ({\cal C}{\cal P}
{\cal T})^2 = -1$. The mass dimensions of these spinors is 1 and the 
Elkos ($\lambda(x)$) Lagrangian is:
{\small
\beq
\label{eq:ElkoLag}
\mathfrak{L}_{elko}= 
\frac{1}{2} 
\l[\frac{1}{2} g^{\mu \nu}(\mathfrak{D}_\mu \blambda \mathfrak{D}_\nu \lambda 
+ \mathfrak{D}_\nu \blambda \mathfrak{D}_\mu \lambda) \r] 
- V(\blambda \lambda) \, .
\eeq
}
where $\lambda(x)$ is the dual and 
$\mathfrak{D}_\nu$ is the covariant derivative \cite{ours}.
%
%\beq
%\mathfrak{D}_\mu \lambda = 
%(\overrightarrow{\partial}_\mu + \Omega_\mu)\,\lambda(x)~;~
%%
%\mathfrak{D}_\mu\blambda = 
%\blambda(x) \, (\overleftarrow{\partial}_\mu - \Omega_\mu) \, ,
%\eeq
%
%$\Omega_{\mu}$ is the tangent space connection and $\gamma$'s are the
%Dirac matrices in the Weyl representation Refs. \cite{Boehmer2007i,ours}.  
%Elko $\lambda(x)$ and its dual $\blambda(x)$ have the following form:
%{\small	
%\begin{equation}
%\label{eq:elkosform}
%\lambda(x) =\left(\begin{array}{c} 
%              \pm \sigma_2 {\phi^{(1)}}^* \\ 
%              \phi^{(1)} 
%              \end{array} \right)
%
%\qquad 
%\blambda(x) = i \left({\phi^{(2)}}^{\dagger} 
%                   \pm {\phi^{(2)}}^{\dagger} \sigma_2 
%             \right)
%\end{equation}
%}
Elkos have mass dimension one, hence, the only
power counting renormalizable interactions of Elkos with standard
matter particle take place through Higgs doublet or  gravity
\cite{Ahluwalia2005}. Elkos are dark matter
candidates \cite{Ahluwalia2005} and are also refereed as dark spinors.

\section{FRW background}

In the case of homogeneous-isotropic FRW background, the Einstein
equations demand that $\lambda$ and $\blambda(x)$ should depend on
single scalar function \cite{Boehmer2007i,ours} 
%
%\beq
$\lambda(x)  = \overline{\varphi}(\eta) \lambda_0;\quad
\blambda(x) = \overline{\varphi}(\eta) \lambda_1
$
%\eeq
where $\lambda_0, \lambda_1$ are constant column and row vectors,
respectively. The acceleration and Friedman equations are then given
by:
 \br 
\label{eq:Eacceleration}
& & {\cal H}' = \frac{1}{3 \MPl} 
\l[a^2(\eta) V(\overline{\varphi}) - {\bphi'}^2(\eta)\r] \\
\label{eq:EFriedmann}
& &  \mathcal{H}^2 = \frac{1}{3 \MPl} 
\l[\frac{{\bphi'}^2/2 
      + a^2(\eta) V(\overline{\varphi})}{1 + \Fpt}\r] 
\qquad  \quad \Fpt = \frac{\bphi^2}{8 \MPl} 
\er
The following points are worth noting regarding the above results:
(i) The background Elko (and its dual) depend on a single scalar
function $(\overline{\varphi})$ satisfying $ \overline{\blambda}
\overline{\lambda} = \overline{\varphi}^2(\eta)$. This can be
interpreted as an Elko-pair forming a scalar
condensate --- spinflaton. 
(ii) The acceleration equation for the spinflaton
(\ref{eq:Eacceleration}) is identical to canonical scalar field
inflation. However, the Friedmann (\ref{eq:EFriedmann}) equations have
non-trivial corrections due to Elko. The Elko modification to the
canonical inflaton equations are determined by $\Fpt$. 
(iii) The form of the potential and the evolution of the field is completely
different compared to the canonical scalar field. 

\section{Linear perturbation}

Let us consider linear perturbation of FRW background in the 
longitudinal gauge \cite{Kodama-Sasa:1984}. As in standard inflation, 
the tensor perturbations do not couple to the energy density and pressure 
inhomogeneities and they are free gravitational waves \cite{Kodama-Sasa:1984}.
%
%\beq
%\label{eq:TensorPertu}
%\mu_{_T}'' + \l(k^2 - \frac{a''(\eta)}{a(\eta)}\r)\mu_{_T} = 0 \, .
%\eeq
%

However, the scalar perturbations are harder to derive; unlike the
scalar field, Elkos are described by four complex functions which 
are related by constraints. The most general perturbed Elko can lead to
the scalar and vector perturbations. Besides, it can also lead to a
non-vanishing anisotropic stress. For simplicity, we will assume that
the anisotropic stress is identically zero.

Perturbation of a spinor has long been studied in spherically
symmetric Skyrmion model using the hedgehog ansatz
\cite{Chodos1975}. We use the similar procedure for the perturbed Elko
and its dual in the case of perturbation theory \cite{ours}. 
As in the canonical scalar field, the perturbation equations can be combined 
in to a single equation in-terms of modified Mukhanov-Sasaki variable, i. e.,
{\small	
\br
\label{eq:MSequation}
& & Q'' - \l[\nabla^2  + \frac{z''}{z} 
        - \ln[1-\Fpteps]'' 
        + \frac{7 \Hm' \Fpteps^{\frac{1}{2}} }{2}
        + \frac{\Hm \varepsilon' 
       \Fpteps^{\frac{1}{2}}}{\varepsilon}\r] Q \simeq 0 \\
& & Q =  a \, \dphi + z \, \Psi;~~ z = \left[1- \Fpteps \right] (a \bphi')/\Hm~;~~
\Fpteps = \frac{\cal F}{{\cal F} + \varepsilon}
\er
}
and $\varepsilon$ is the slow-roll parameter.
Invoking the slow-roll conditions ($\varepsilon, |\delta| \ll 1$) in
the scalar (\ref{eq:MSequation}) and tensor perturbation equations
 --- and following the standard quantization
procedure  --- the scalar and tensor
power spectra are given by:
{\small
\beq
\label{eq:SPS}
{\cal P}_S(k) \simeq \l(\frac{H^2}{8 \MPl \pi^2}\r)
\left(\frac{\varepsilon + \Fpt}{\varepsilon^2}\right)~;~
%\l[1- 2(c_0+1)\varepsilon_{_{\rm can}} \r] \\
%
%\label{eq:TPS}
{\cal P}_T(k) \simeq \l(\frac{2 H^2}{\MPl \pi^2}\r) 
%\l[1- 2(c_0+1) \varepsilon_{_{\rm can}} + 2 \varepsilon_{_{\rm can}} x\r]
\eeq
}
\hspace*{-7pt} 
%where $c_0=\gamma_{_{\rm Euler}}+ \ln2- 2$ is a constant, $x=\ln(k^*/k)$
%and $k^*$ is the pivot scale.

The above spectra allow us to draw important conclusions. Firstly,
the scalar and tensor power spectra of the spinflaton, in slow-roll,
are nearly scale-invariant \cite{ours}. Secondly, in the leading order
of $\epsilon_{_{\rm can}}$, the spectra has a non-zero running:
{\small
\begin{subequations}
%\numparts
\br
\frac{d n_S}{d \ln{k}} &=&
-\frac{\varepsilon_{_{\rm can}}}{2}- 4 \varepsilon_{_{\rm can}}
\Fpteps^{1/2} + \frac{\varepsilon_{_{\rm can}}}{2}\frac{\Fpt}{1+\Fpt} 
\er
\end{subequations}
}
\hspace*{-7pt} It is interesting to note that the running of scalar spectral 
index ($-0.09 < d n_S/d (\ln k) < 0.019$) is consistent with the
WMAP-5 year results \cite{Komatsu2008}. For instance, using the WMAP
value of $\varepsilon_{_{\rm can}} = 0.038$ \cite{Komatsu2008} and
assuming that $\Fpt$ is tiny, we get $d n_S/d (\ln k) \sim
-0.019$. This is the one of the main predictions of spinflation.

Thirdly, the tensor-to-scalar ratio $r$ is no longer equal to $16
\varepsilon_{_{\rm can}}$ and is given by:
\beq
\label{eq:r}
r \simeq 16 \, \varepsilon_{_{\rm can}} \, \l[1- 2 \Fpteps\r] \, .
\eeq
Physically, this suggests that the gravitational wave contribution
during slow-roll spinflation is smaller than for canonical
inflation. Lastly, as for the canonical scalar field, the scalar and
tensor perturbations during spinflation originate from the scalar
condensate and they are not independent. Hence, consistency relations
link them together. The one which is observationally useful is the
relation between $n_{_T}$ and $r$:
{\small
\beq
n_{_T}=\frac{r}{8} (1+\Fpteps) \l[ 1  
               + \varepsilon_{_{\rm can}}\l[\frac{11}{6} c + \Fpteps-\Fpt \r] 
               - 2\delta_{_{\rm can}} \, c \r] \, .
\eeq
}
\indent To conclude, we have shown that the spinor condensate 
in the early universe is a viable model of cosmological inflation.  It
leads to the identical acceleration equation as that of canonical
scalar field driven inflation. We have used the Hedgehog ansatz to
obtain the scalar perturbation and, in the slow-roll limit, we have
shown that scalar and tensor perturbations are nearly scale
invariant. The model predicts a running of scalar spectral index
consistent with WMAP-5 year data. The consistency relation between the
scalar and tensor spectra are non-trivial and have different feature
compared to the models where the scalar fields are considered
elementary.

The author wishes to thank Damien Gredat, Christian Boehmer and Roy
Maartens for discussions. The work is supported by the Marie Curie
Incoming International Grant IIF-2006-039205.

%%%%%%%%%%%%%%%%%%%%%%%%%%%%%
%%%%%%% bibliography
%%%%%%%%%%%%%%%%%%%%%%%%%

%\input{Mar11-Ver.bbl}


\begin{thebibliography}{22}
\expandafter\ifx\csname natexlab\endcsname\relax\def\natexlab#1{#1}\fi
\expandafter\ifx\csname bibnamefont\endcsname\relax
  \def\bibnamefont#1{#1}\fi
\expandafter\ifx\csname bibfnamefont\endcsname\relax
  \def\bibfnamefont#1{#1}\fi
\expandafter\ifx\csname citenamefont\endcsname\relax
  \def\citenamefont#1{#1}\fi
\expandafter\ifx\csname url\endcsname\relax
  \def\url#1{\texttt{#1}}\fi
\expandafter\ifx\csname urlprefix\endcsname\relax\def\urlprefix{URL }\fi
\providecommand{\bibinfo}[2]{#2}
\providecommand{\eprint}[2][]{\url{#2}}

\bibitem{Komatsu2008}
\bibinfo{author}{\bibfnamefont{E.}~\bibnamefont{Komatsu}} \bibnamefont{et~al.}
  (\bibinfo{collaboration}{WMAP}) (\bibinfo{year}{2008}), \eprint{0803.0547}.


%\bibitem{Starobinsky1980}
%\bibinfo{author}{\bibfnamefont{A.~A.} \bibnamefont{Starobinsky}},
%  \bibinfo{journal}{Phys. Lett.} \textbf{\bibinfo{volume}{B91}},
%  \bibinfo{pages}{99} (\bibinfo{year}{1980}).

\bibitem{Ford1989}
\bibinfo{author}{\bibfnamefont{L.~H.} \bibnamefont{Ford}},
  \bibinfo{journal}{Phys. Rev.} \textbf{\bibinfo{volume}{D40}},
  \bibinfo{pages}{967} (\bibinfo{year}{1989}); 
%\bibitem[{\citenamefont{Golovnev et~al.}(2008)\citenamefont{Golovnev, Mukhanov,
%  and Vanchurin}}]{Golovnev2008}
\bibinfo{author}{\bibfnamefont{A.}~\bibnamefont{Golovnev}}
et al,
%  \bibinfo{author}{\bibfnamefont{V.}~\bibnamefont{Mukhanov}}, \bibnamefont{and}
%  \bibinfo{author}{\bibfnamefont{V.}~\bibnamefont{Vanchurin}},
  \bibinfo{journal}{JCAP} \textbf{\bibinfo{volume}{0806}}, \bibinfo{pages}{009}
  (\bibinfo{year}{2008});
%\bibitem[{\citenamefont{Parker and Zhang}(1993)}]{Parker1993}
%\bibinfo{author}{\bibfnamefont{L.}~\bibnamefont{Parker}} \bibnamefont{and}
%  \bibinfo{author}{\bibfnamefont{Y.}~\bibnamefont{Zhang}},
%  \bibinfo{journal}{Phys. Rev.} \textbf{\bibinfo{volume}{D47}},
%  \bibinfo{pages}{416} (\bibinfo{year}{1993});
K.~Dimopoulos, D.~H.~Lyth and Y.~Rodriguez,
  %``Statistical anisotropy of the curvature perturbation from vector field
  %perturbations,''
  arXiv:0809.1055.

%\bibitem{Shafi:1986vv}
%  Q.~Shafi and C.~Wetterich,
  %``INFLATION FROM HIGHER DIMENSIONS,''
%  Nucl.\ Phys.\  B {\bf 289}, 787 (1987).
  %%CITATION = NUPHA,B289,787;%

\bibitem{Boehmer2007i}
\bibinfo{author}{\bibfnamefont{C.~G.} \bibnamefont{Boehmer}},
  \bibinfo{journal}{Annalen Phys.} \textbf{\bibinfo{volume}{16}},
  \bibinfo{pages}{325} (\bibinfo{year}{2007});
%, \eprint{gr-qc/0701087}.
%\bibitem{Boehmer2008}
%\bibinfo{author}{\bibfnamefont{C.~G.} \bibnamefont{Boehmer}},
Phys.\ Rev.\  D {\bf 77} (2008) 123535,
%  (\bibinfo{year}{2008}), 
\eprint{0804.0616};
%\bibitem[{\citenamefont{Boehmer and Mota}(2008)}]{Boehmer2008c}
\bibinfo{author}{\bibfnamefont{C.~G.} \bibnamefont{Boehmer}} \bibnamefont{and}
  \bibinfo{author}{\bibfnamefont{D.~F.} \bibnamefont{Mota}},
  \bibinfo{journal}{Phys. Lett.} \textbf{\bibinfo{volume}{B663}},
  \bibinfo{pages}{168} (\bibinfo{year}{2008}), \eprint{0710.2003}.

%\bibitem{TC} 
%C. J. Pethick and H. Smith, {\sl Bose-Einstein Condensation in 
%Dilute Gases}, Cambridge University Press (2002)

%\bibitem{weinberg}
%S. Weinberg, {\sl The quantum theory of Fields, Vol. 1: Foundations},
%Cambridge University Press (1995).

\bibitem{Ahluwalia2005}
D.~V.~Ahluwalia et al,
 % \bibnamefont{\&}
 % \bibinfo{author}{\bibfnamefont{D.}~\bibnamefont{Grumiller}},
  \bibinfo{journal}{Phys. Rev.} \textbf{\bibinfo{volume}{D72}},
  \bibinfo{pages}{067701} (\bibinfo{year}{2005});  
\bibinfo{journal}{JCAP} \textbf{\bibinfo{volume}{07}}, \bibinfo{pages}{012}
  (\bibinfo{year}{2005}).%, \eprint{hep-th/0412080}.

%\bibitem{Ahluwalia:2008xi}
%  D.~V.~Ahluwalia et al,  arXiv:0804.1854 [hep-th].
  %%CITATION = ARXIV:0804.1854;%%

%\bibitem{Streater-Wighman}
%R. F. Streater, A. S. Wightman, 
%{\sl PCT, spin and statistics and all that} (Benjamin, 1964)

\bibitem{ours}
 D.~Gredat and S.~Shankaranarayanan, JCAP 01, 008 (2010)
  %``Consistency relation between the scalar and tensor spectra in
  %spinflation,''
  arXiv:0807.3336 [astro-ph]; S.~Shankaranarayanan, To appear in 
Int. J. Mod. Phys. D., arXiv:0905.2573

\bibitem{Kodama-Sasa:1984}
\bibinfo{author}{\bibfnamefont{H.}~\bibnamefont{Kodama}} \bibnamefont{and}
  \bibinfo{author}{\bibfnamefont{M.}~\bibnamefont{Sasaki}},
  \bibinfo{journal}{Prog. Theor. Phys. Suppl.} \textbf{\bibinfo{volume}{78}},
  \bibinfo{pages}{1} (\bibinfo{year}{1984});
%\bibitem[{\citenamefont{Mukhanov et~al.}(1992)\citenamefont{Mukhanov, Feldman,
%  and Brandenberger}}]{Mukhanov-Feld:1992}
\bibinfo{author}{\bibfnamefont{V.~F.} \bibnamefont{Mukhanov}},
  \bibinfo{author}{\bibfnamefont{H.~A.} \bibnamefont{Feldman}},
  \bibnamefont{and} \bibinfo{author}{\bibfnamefont{R.~H.}
  \bibnamefont{Brandenberger}}, \bibinfo{journal}{Phys. Rept.}
  \textbf{\bibinfo{volume}{215}}, \bibinfo{pages}{203} (\bibinfo{year}{1992}).


\bibitem{Chodos1975}
\bibinfo{author}{\bibfnamefont{A.}~\bibnamefont{Chodos}} \bibnamefont{and}
  \bibinfo{author}{\bibfnamefont{C.~B.} \bibnamefont{Thorn}},
  \bibinfo{journal}{Phys. Rev.} \textbf{\bibinfo{volume}{D12}},
  \bibinfo{pages}{2733} (\bibinfo{year}{1975});
%\bibitem[{\citenamefont{Deser et~al.}(1976)\citenamefont{Deser, Duff, and
%  Isham}}]{Deser1976a}
\bibinfo{author}{\bibfnamefont{S.}~\bibnamefont{Deser}}
et al,
%  \bibinfo{author}{\bibfnamefont{M.~J.} \bibnamefont{Duff}}, \bibnamefont{and}
%  \bibinfo{author}{\bibfnamefont{C.~J.} \bibnamefont{Isham}},
  \bibinfo{journal}{Nucl. Phys.} \textbf{\bibinfo{volume}{B114}},
  \bibinfo{pages}{29} (\bibinfo{year}{1976}).

\end{thebibliography}
\end{document}